\documentclass[useAMS,usenatbib]{mn2e}
\usepackage{graphicx}
\usepackage{txfonts}

\catcode`\@=11
\def\gsim{\ifmmode{\mathrel{\mathpalette\@versim>}}
    \else{$\mathrel{\mathpalette\@versim>}$}\fi}
\def\lsim{\ifmmode{\mathrel{\mathpalette\@versim<}}
    \else{$\mathrel{\mathpalette\@versim<}$}\fi}
\def\@versim#1#2{\lower 2.9truept \vbox{\baselineskip 0pt \lineskip
    0.5truept \ialign{$\m@th#1\hfil##\hfil$\crcr#2\crcr\sim\crcr}}}
\catcode`\@=12
\newcommand{\msun}{\ensuremath{\rm M_\odot}}

\def\msun{\hbox{$M_\odot$}}

\def\yr-1{\hbox{${\rm yr}^{-1}$}}
\begin{document}
\title{A Different Approach to Galaxy Evolution}
\author[Alvio Renzini]{Alvio Renzini$^{1}$\thanks{E-mail: 
alvio.renzini@oapd.inaf.it}\\ 
 $^{1}$INAF - Osservatorio
Astronomico di Padova, Vicolo dell'Osservatorio 5, I-35122 Padova,
Italy}

\date{Accepted June 25, 2009; Received May 6, 2009 in original form}
 \pagerange{\pageref{firstpage}--\pageref{lastpage}} \pubyear{2002}

\maketitle
                                                            
\label{firstpage}

\begin{abstract}
The consequences are explored of an observationally established
relation of the star formation rate (SFR) of star-forming galaxies
with their stellar mass ($M$) and cosmic time ($t$), such that SFR
$\propto M\, t^{-2.5}.$ It is shown, that small systematic differences in SFR
dramatically amplify in the course of time: galaxies with above
average SFR run into quasi-exponential mass and SFR growth, while
galaxies with below average SFR avoid such exponential growth and
evolve with moderate mass increase. It is argued that galaxies
following the first path would enormously overgrow if keeping to form
stars all the way to the present, hence should quench star formation
and turn passive. By the same token, those instead avoiding the
quasi-exponential growth may keep to form stars up to the present.
Thus, it is conjectured that this divergent behaviour can help
understanding the origin of the dichotomy between passive, spheroidal
galaxies, and star-forming, disk galaxies.

\end{abstract}



\maketitle


\section{Introduction}
In this paper I propose a different approach towards understanding galaxy
evolution.  Instead of starting from first physical principles and
proceed deductively, as in the widely explored $\Lambda$CDM approach,
I will attempt a fully {\it inductive}, bottom-up  approach based exclusively 
on few established empirical evidences.

Indeed, in recent years a formidable body of multiwavelength data have
been accumulated on galaxies at all redshifts up to $\sim 6$. Such
data are especially extended for $z\lsim 3$, hence encompassing the
major epoch of galaxy growth peaking at $z\sim 2$, when the
morphological differentiation into (spiral) disks and (elliptical)
spheroids is well under way. Various multiwavelength photometric and
spectroscopic databases have allowed several groups to derive major
galaxy quantities such as redshifts, star formation rates (SFR),
stellar masses ($M$), etc.

One important result of these observational studies, based on the
GOODS database (Giavalisco et al. 2004; Vanzella et al. 2008, and
references therein) was the recognition that at $1.4\lsim z\lsim 2.5$
the SFR of star-forming (SF) galaxies tightly correlates with the stellar
mass, with SFR $\propto\sim M$, while some galaxies have already
ceased to form stars and evolve passively (Daddi et al. 2007). As
illustrated in Figure 1, at these redshifts galaxies are either
actively star forming, or already passive, with very few galaxies
lying out of these two main branches, i.e., the active branch with SFR
$\propto\sim M$, and the passive one with SFR $\sim 0$. These
evidences lead to recognize that the vast majority of the SF galaxies
are not in a starburst phase, even if their SFRs are hundreds of
$\msun\yr-1$. Instead, they are steadily forming stars at high rates
over a major fraction of the $\sim 2$ Gyr of cosmic time from $z\sim
3$ to $z \sim 1.4$.

\begin{figure}
\includegraphics[width=84mm]{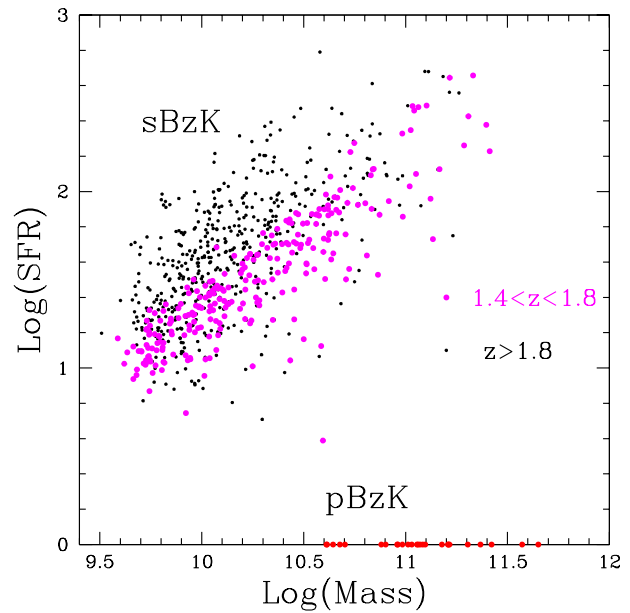}
\caption{The star formation rate vs. stellar mass (respectively in
$\msun\yr-1$ and $\msun$) for $BzK$-selected galaxies at $1.4<
z < 2.5$ in the GOODS-South field (from Daddi et al. 2007). The SFR
was derived form the UV flux plus extinction correction. Galaxies
cluster either on the star-forming branch (sBzK) or on the passive 
branch (pBzK) with star formation rate SFR$\simeq 0$, displayed here 
at Log(SFR)=0. Notice that very few galaxies lie in between the two
main branches, indicative of SF quenching being a fast process.}
\label{goods}
\end{figure}

The same pattern, with galaxies lying either on a tight SF branch with
SFR $\propto\sim M$ or being already passive, has then been recognized
also at lower redshifts, from $z\sim 1$ (Elbaz et al. 2007), all the
way to $z\sim 0$, hence revealing a steady decrease of the SFR of
galaxies in the SF branch, which at fixed mass is by a factor $\sim
30$ between $z\sim 2$ and $z\sim 0$ (Daddi et al. 2007, see their
Fig. 17).  Due to its tightness, the SF branch has been dubbed the
{\it main sequence} of SF galaxies (Noeske et al. 2007).

In Daddi et al. (2007) the SFRs of $1.4\lsim z\lsim 2.5$ galaxies
have been derived with the traditional method based on the rest-frame
UV flux, plus extinction correction from the slope of the UV continuum.
SFRs in full agreement with these UV-derived ones have been recently
obtained from stacking the 1.4 GHz fluxes of $\sim 12,000$ SF galaxies
in the COSMOS field (Pannella et al. 2009), hence confirming the
reliability of the classical procedure for the extinction correction. In
particular, an almost perfect linear relation of the SFR with $M$ is
recovered, hence implying a specific SFR (SSFR= SFR$/M$) independent
of mass. As in Daddi et al. (2007), $1.4\lsim z\lsim 2.5$ galaxies
were selected using the $BzK$ criterion introduced by Daddi et
al. (2004), and applying it to the deep $K$-band selected catalog of
galaxies in  the COSMOS field (McCracken et al. 2009).

Combining their own data for galaxies at $1.4\lsim z\lsim 3$ with
those at $z\sim 1$ (Elbaz et al. 2007), $z\sim 07$ (Noeske et
al. 2007) and $z\sim 0$ (Brinkmann et al. 2004), Pannella et al
(2009) have then obtained the following best-fit relation for galaxies
on the SF branch:

\begin{equation}
<\!{\rm SFR}\!> \simeq 270\times \eta M_{11} (t/3.4\times 10^9)^{-2.5}\quad
            \quad (\msun\yr-1),
\end{equation}
where $M_{11}$ is the stellar mass in units of $10^{11}\,\msun$, and
$t$ is the cosmic time in years. The factor $\eta={\rm
SFR}(t=3.4\times 10^9)/270$ has been introduced here, with $\eta\equiv
1$ for the best fit relation presented by Pannella et al. (2009), but
the effects of assuming other values will be also investigated. 
Note that a quite similar relation, consistent with Eq. (1), can be
derived from Fig. 17 in Daddi et al. (2007).

SFRs and stellar masses used in establishing Eq. (1) were obtained
assuming a Salpeter (1955) initial mass function (IMF). Adopting other
IMFs such as those of Kroupa (2001) or Chabrier (2003) would affect
SFRs and masses by the same factor, thus leaving SSFRs
unchanged. Therefore, none of the results presented in this paper
depends on the adopted IMF, provided it does not vary from one galaxy
to another (e.g., as a function of mass), or from one cosmic time to
another. If this were the case, the net effect would be to change the
exponents of $M$ and/or $t$ in Eq. (1), a possibility that is not
further explored in this paper given the unconstrained nature that
such an exercise would have. 

I now explore the consequences of assuming that Eq. (1)
adequately describes the evolution of the SFR for galaxies in
the SF branch from $z\sim 3$ ($t\sim 2$ Gyr) all the way to $z\sim 0$
($t\sim 13.7$ Gyr).

Before exploiting this relation, it is worth mentioning that other
studies find SSFRs markedly declining with mass (e.g., Erb et
al. 2006; Cowie \& Barger 2008). Dunne et al. (2008) have discussed
this kind of discrepancies and their possible origin, and confirm a
SSFR nearly independent of mass at $z\sim 2$, as most recently
found by Santini et al. (2009) for galaxies in the GOODS-South
field. Notice also that Noeske et al. (2007) find SFR $\propto\sim
M^{2/3}$ for $z\lsim 1$ galaxies on the SF branch. Hence the exponent
of $M_{11}$ in Eq. (1) may not be strictly 1, and it may slightly
evolve with redshift (cf. Dunne et al. 2008).  Still, for the present
exercise I explore the implications of Eq. (1) taking it at face value
in the redshift interval $0<z<3$.

\section{The Growth of Galaxies}

Eq. (1) also describes as the stellar mass of individual galaxies grows
as a result of their star formation, and does so as a function of
stellar mass and time. Hence, one can integrate the equation
$dM/dt\;=\;{\rm SFR}$, with SFR given by the right hand side of
Eq. (1). This leads to a galaxy growth with time described by the
equation:
\begin{equation}
{M(t)\over M(2\;{\rm Gyr})}\simeq {\rm exp}(13.53\;\eta)\,
              {\rm exp}(-38.26\;\eta\; t^{-1.5}),
\end{equation}
which represents the growth factor of galaxy mass as a function
of cosmic time.  Notice that it is not attempted here to explore
galaxy evolution beyond $z=3$ ($t\lsim 2$ Gyr), as Eq. (1) is
observationally established only for $z\lsim 3$, but just to follow
the mass growth from $t\sim 2$ Gyr onwards.  Combining Eq. (1) and (2)
one then derives the corresponding evolution with time of the SFR of
individual galaxies:
\begin{equation}
{{\rm SFR}(t)\over{\rm SFR}(2\; {\rm Gyr})}=5.66\times 
          {M(t)\over M(2\;{\rm Gyr})}\times t^{-2.5}.
\end{equation}

One intriguing aspect of the SFR as given by Eq. (1) is that its
normalization $\eta$ appears at the exponential in Eq. (2) and
(3). Hence, the effects of relatively small differences in $\eta$ dramatically
amplify as time goes by. This is illustrated in Fig. 2 where the
cases with $\eta=1$, 1/2 and 1/4 are shown. Let us first focus on
the $\eta=1$ case. Notice the extremely rapid growth of the stellar
mass, amounting to more than a factor $\sim 10^5$,
if Eq. (1) were to hold true from $t=2$ Gyr to the present (from
$z\sim 3$ to $z=0$). Clearly, observations do not support such a
dramatic overgrowth. However, with $\eta =1/4$, i.e., just a factor of 4
lower SFR for given mass and time, the corresponding growth is much
smaller, i.e., just a factor $\sim 30$.

The parameter $\eta$ is meant to describe two independent aspects of
Eq. (1): (i) exploring the effects of a possible systematic offset of
the derived SFRs, which certainly cannot be currently excluded, and
(ii) explore the effect of departures of the SFRs of individual
galaxies from the average, i.e., for being systematically higher/lower
than the average by a factor $\eta$. For the first aspect, Fig. 2
shows that the true value of the SSFR at given time critically
determines the subsequent grow rate of the stellar mass of galaxies: a
factor of just a few difference making enormous differences in the
subsequent evolution. This means that the average SSFR would need to
be measured with extreme accuracy in order to accurately predict such
a subsequent evolution. Current systematic SFR uncertainties are
indeed by a factor of $\sim 2$ or 3, hence $\eta$ in Eq. (1) will have
to be used as an adjustable parameters within these observational
uncertainties.

The second aspect is perhaps more attractive. It implies that at given
mass and cosmic time galaxies whose SSFR differ by a relatively small
factor experience radically different mass evolutions: some enjoy a
rather modest mass growth, with secularly declining SFRs, while others
suffer a runaway, quasi-exponential mass growth, which certainly
cannot be sustained for more than $\sim 1$ Gyr. Eq. (1) refers to the
{\it average} SFR, hence quite naturally one expects some galaxies to
have SFRs systematically lower than the average, and others to have it
higher than the average. However small this dispersion can be, it
naturally tends to dramatically amplify in the course of time, as
demanded by Eq.s (2) and (3) and illustrated in Fig. 2 and
Fig. 3.

\begin{figure}
\includegraphics[width=84mm]{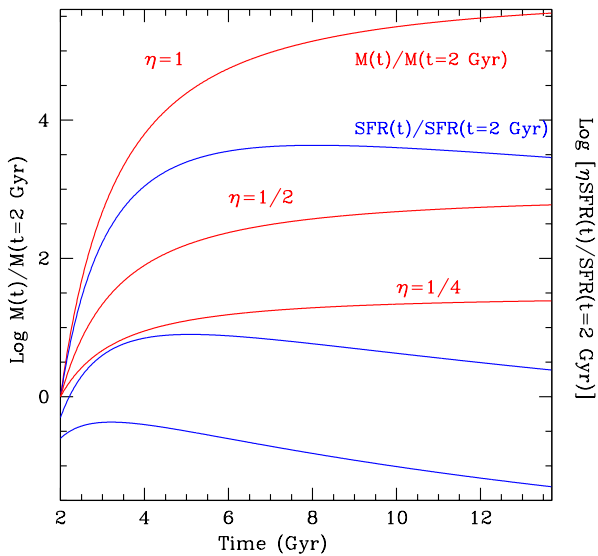}
\caption{The growth with cosmic time of the stellar mass normalised at
its initial value at $t=2$ Gyr ($z\sim 3$), following Eq. (1) and (2),
and for three values of $\eta$ as indicated. Also shown is the
corresponding evolution with time of the SFR, following Eq. (3), for the same
values of $\eta$. The three curves are initially offset by a factor
$\eta$ to show the initial difference in their SFRs (i.e., at $t=2$
Gyr). One can appreciate that SFRs for given mass and time that differ
by only a factor of a few lead to vastly different evolutionary paths.}
\label{growth}
\end{figure}

\begin{figure}
\includegraphics[width=84mm]{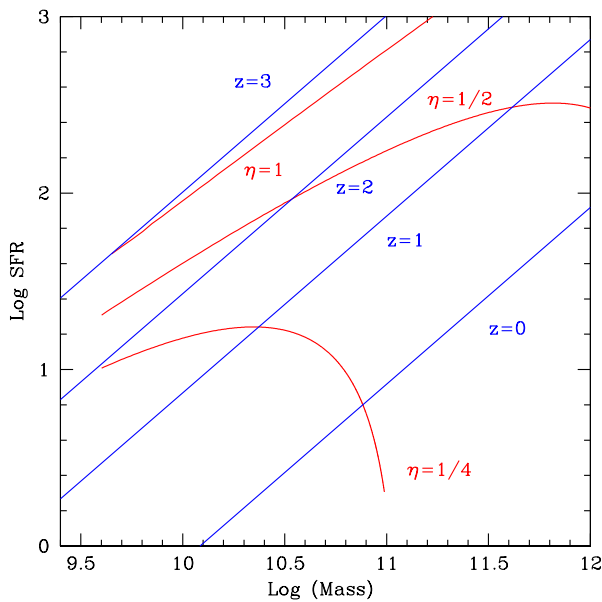} 
\caption{The evolution of stellar mass and SFR for galaxies which at
$t=2$ Gyr start with $M=4\times 10^{10}\,\msun$, for the three different
values of $\eta$ also used for Fig. (1). The straight lines show the SFR-mass 
relation from Eq. (1) at four different redshifts and with $\eta=1$.
}
\label{growth}
\end{figure}

The likely origin of such a dispersion is environment. As mentioned
above, the tight SF branch of galaxies indicates that they experience
(quasi-)steady SF. This picture is in agreement with recent hydrodynamical
simulations in which SF in galaxies is continuously fed by {\it cold
stream} gas accretion from the environment (Dekel et
al. 2009). Therefore, galaxies in different environments are likely to
experience different rates of gas accretion, hence different
SSFRs. Actually, Eq. (1), in spite of its simplicity, may capture both
{\it nature} and {\it nurture} aspects of galaxy evolution, which to
some extent undoubtedly must co-exist. Indeed, the stellar mass,
certainly a main driver in galaxy evolution, clearly stands for {\it
nature}, and a dispersion of $\eta$ results from a dispersion in the
physical properties of the local environment of individual galaxies
({\it nurture}). Moreover, the $t^{-2.5}$ factor in Eq. (1) describes
the global, cosmological evolution of the environment, a combination
of cosmic expansion and the progressive consumption of the cold-gas
reservoir, as more baryons are shock heated to virial temperatures, or
even above it by feedback effects (galactic winds).

Notice that in galaxies undergoing rapid mass accretion ($\eta\gsim
1$) the SFR increases quasi-exponentially with time, i.e., just the
opposite of what assumed in the so-called $\tau$-models in which SFR
{\it decreases} exponentially with time. The unfitness of such models
to describe some major aspects galaxy evolution was pointed out in
Cimatti et al. (2008) and is further explored in Maraston et
al. (2009, in preparation).

\section{A Conjecture on the Origin of Morphological Differentiation}

The origin of the sharp separation into the early-type (spheroid) and
late-type (disk) families of galaxies remains a central question in
galaxy evolution studies. Based on the above arguments, I would like
to propose here a conjecture that may help understanding the
origin of this dichotomy. I assume that Eq. (1) for the average SFR
holds true for SF galaxies with $\eta=1$ (but a slightly lower value
of $\eta$ may work even better, see below). Then individual galaxies
evolve according to Eq.s (2) and (3), each with its specific value of
$\eta$, with a dispersion of $\eta$ values similar to the empirical
dispersion of the SFRs shown in Fig. 1. In practice, a range of
$\eta$ within a factor $3-4$ about its mean value should encompass
the vast majority of galaxies in the SF branch.

As shown in Fig. 2, galaxies with $\eta\gsim 1$ undergo extremely fast
mass growth that cannot be sustained indefinitely. At some point in
time a SFR as given by the r.h.s. of Eq. (1) cannot apply any longer,
which is to say galaxies must leave the SF branch described by
Eq. (1). As suggested by Pannella et al. (2009), the only way this can
happen is by quenching completely SF, thereby galaxies join the
passive branch (indeed, they have no other place to go in Fig. 1).
This SF quenching can happen in a variety of ways. Just to mention
one, extremely high gas accretion by, and SF rates in, massive disks
at $z\sim 2$ likely results in disk instabilities, with massive clumps
coalescing at the center to form a bulge, feeding an AGN and its
ensuing feedback (Immeli et al. 2004; Elmegreen \& Elmegreen 2006;
Genzel et al. 2008).

Galaxies with sub-average gas accretion and SFR ($\eta\lsim 1$), on
the contrary, avoid the quasi-exponential mass and SFR growth; their
mass increases moderately and they exhibit a slowly decreasing SFR
over most of the cosmic time (Fig. 2 and Fig. 3). Disk galaxies with
sub-average SFRs are therefore likely to avoid global disk
instabilities, hence retaining their disk structure all the way to the
present.

Notice also that those galaxies that experience a
quasi-exponential growth naturally develop an $\alpha$-element
enhancement, that is typical of ellipticals and bulges (e.g., Thomas
et al. 2005; Zoccali et al. 2006). Instead, those galaxies that avoid
a quasi-exponential growth experience a chemical enrichment to which
both supernova types contribute substantially, hence resulting in
near-solar abundance ratios.

In summary, the tenet of the conjecture is that the morphological
differentiation of galaxies is the result of a SSFR (almost) independent of
mass working as a very efficient amplifier of galaxy to galaxy
differences of SSFR.  Galaxies with above-average SFRs experience a
runaway mass accretion resulting in global instabilities and
spheroid formation. Instead, those with sub-average SFRs experience
only a modest mass growth, avoid instabilities, and survive as disks.
Differences in SSFR likely arise from differences in cold gas
accretion from the environment, which can also help understanding the
origin of the morphology-density relation.

\section{Caveats}

This scenario completely neglects mergers, and assumes {\it in situ}
SF as the only process leading to the growth of galaxy mass.  In
recent years there has been a marked shift of emphasis from (major)
mergers to cold stream accretion as the main driver of galaxy
evolution (Genzel et al. 2008; Dekel et al. 2009, and references there
in). Even so, mergers must occur and play a role that may indeed be
dominant at very high redshifts (say, $z\gsim 3$), but then steadily
decline (e.g., Masjedi, Hogg \& Blanton 2008; Conselice, Yang \& Bluck
2009) and is superseded by cold stream accretion (Dekel et
al. 2009). In any event, a full description of galaxy evolution must
also include merging processes. Like done here for star formation
alone, one could include this effect using empirical merger rates once
they are firmly established.

In the simplified approach presented here, it is assumed that SF
galaxies evolve following Eq.s (2) and (3), each with a fixed value of
$\eta$. Actually, gas accretion and ensuing SFR must fluctuate up and
down as a function of time, an aspect that indeed may be regarded as a
series of {\it minor merger} events (Dekel et al. 2009). So, the
evolution of individual galaxies cannot be so smooth as implied by
Eq.s (2) and (3) and shown in Fig. 2. Yet, apart from short timescale
fluctuation, one should expect that different galaxies (in different
environments) experience systematically higher/lower than average gas
accretion and SF rates, once averaged over sufficiently long
timescales.  It is indeed this kind of {\it noise suppressed}
evolution that is described by Eq.s (2) and (3).

On the other hand, as clear from Fig. 2, what matters most is the
value of $\eta$ during the relatively short interval of cosmic time
($2\lsim t\lsim 4$ Gyr) when the quasi-exponential growth may or may not
take place. Later, the SFR tends to decrease (and the mass growth to
slow down) no matter what the value of $\eta$ is, as the factor
$t^{-2.5}$ begins to dominate. Actually, it is unlikely that
environmental effects on SFRs maintain the same direction at all
redshifts. For example, overdensity may promote higher SFRs at high
$z$ when cold gas is more abundant, but at low $z$ overdense regions
such as clusters may become detrimental to SF, as most gas has been
shock-heated to high temperatures within the cluster potential well.
Thus, typical values of $\eta$ are likely to depend on a non separable
combination of overdensity and cosmic epoch.

\section{Perspectives}

As the mapping of the galaxy populations at high redshifts progress,
along with that of their large scale structure distribution, it
becomes increasingly urgent to our curiosity trying to understand what
galaxies at some high redshift become at another, lower redshift.  For
example, whether some SF disks in a certain environment are more
likely to remain SF disks, or will suffer a major, catastrophic event
turning them into passive spheroids. The conjecture presented here may
help identifying one of the major mechanisms driving galaxy evolution,
including its bifurcation into SF disks and passive spheroids.
Yet, certainly many critical issues remain open.

First, nothing is said here on the evolution prior to $t=2$ Gyr,
i.e. on how galaxy form and grow during the first 2 billion years of
cosmic evolution. Available data at $z>3$ are presently
insufficient to attempt an empirical approach similar to that followed
here at lower redshifts.

Assuming an empirically motivated stellar mass function
for galaxies at $z=3$, Eq. s (1), (2) and (3) can in principle be used
to evolve such mass function to lower redshifts. Such an evolution
would be critically dependent on several assumptions, worth mentioning
and discussing them here:

1) The average value of $\eta$, i.e., the absolute normalization of
the SSFR in Eq. (1). All estimates, including those of Daddi et
al. (2007) or Pannella et al. (2009) adopted here, are certainly
affected by a systematic error, hard to pinpoint from observations. As
alluded above, one can suspect that an average $\eta$ somewhat less
than 1 (e.g., $\eta\sim 1/2$) may give a more realistic share
between galaxies running into catastrophic growth and those evolving
more peacefully.  Critical to emphasize here is the important role
played by such a normalization. 

2) The dispersion of $\eta$ values, which along with the average
   $\eta$ concur in determining the evolution of the mass function.

3) The SF quenching mechanism, and its dependence on galaxy mass,
environment, and cosmic epoch. We empirically know, from evidences at
low (Thomas et al.  2005) as well as high redshifts (e.g., Cimatti,
Daddi \& Renzini 2006; Renzini 2006; Bundy et al. 2006; Cimatti
et al. 2008), that massive galaxies are the first to turn passive,
around $z\sim 2$. Then, as time goes by, a fraction of galaxies of
lower and lower masses cease to form stars, while others maintain such
activity all the way to the present. This mass phasing of the SF
quenching process is not a natural consequence of the conjecture
presented in Section 3, and requires additional physics besides the
mass-dependent SFR given by Eq. (1). Indeed, a SSFR for actively
  SF galaxies that is independent of mass inherently does not include
  a {\it downsizing} effect, as all masses grow at the same relative
  rate (see Pannella et al. 2009). Hence, downsizing in SF quenching
  must involve physical phenomena that are not described by a
  SSFR($M,t)$ relation for SF galaxies.

4) Assuming that minor, gas-rich mergers are automatically included in
Eq. (1), the effects of major mergers are left out by such an
approach, a limitations that could again be alleviated with either
empirically or theoretically motivated mergers rates.

In conclusion, playing with these assumptions and exploring the
parameter space may in the future help our understanding of galaxy
evolution. For the time being, I just wish to emphasize that an
empirical relation between SFR, stellar mass, and cosmic time
naturally predicts an extreme amplification of small differences in
SFR during their major epoch of SF at $z\sim 2$.

\section*{Acknowledgments}

I wish to thank Andrea Cimatti and Emanuele Daddi for the many
stimulating discussions on galaxy evolution and high-redshifts galaxy
surveys, and Laura Greggio and Mauro Giavalisco for their critical and
constructive reading of a draft of this paper. The GOODS and COSMOS
Teams are collectively acknowledged for the monumental databases they
have produced on high-redshift galaxies. I also like to acknowledge
the kind hospitality of the Osservatorio Astronomico di Padova, and
INAF for financial support via a ``PRIN'' 2007 entitled
``VVDS/zCOSMOS: Measuring the Joint Evolution of Galaxies and the
Large-Scale Structure of the Universe'' (PI Gianni Zamorani).


\label{lastpage}

\end{document}